\documentclass[letterpaper,10pt]{article} 

\usepackage{opticameet3} 
\usepackage{tabularx}
\usepackage{booktabs}
\usepackage{makecell}
\usepackage{array}
\usepackage{graphicx}
\usepackage{caption}
\usepackage{subcaption}
\usepackage{float}
\usepackage{color,soul}
\usepackage{graphicx, url}
\newcommand{\aj}{\emph{AJ}}                        

\newcommand{\apjl}{\emph{ApJL}}                    
\newcommand{\aap}{\emph{A\&A}}                     
\newcommand{\mnras}{\emph{MNRAS}}                  

\newcommand{\pasp}{\emph{PASP}}                    
\newcommand{\Science}{\emph{Science}}              

\newcommand\authormark[1]{\textsuperscript{#1}}

\usepackage{amsmath,amssymb}
\usepackage[colorlinks=true,bookmarks=false,citecolor=blue,urlcolor=blue]{hyperref} 

\begin{document}
\pagestyle{plain}
\pagenumbering{arabic}

\title{Machine Learning for Exoplanet Detection:\\
A Comparative Analysis Using Kepler Data}


\author{Reihaneh Karimi\authormark{1*}, Mahdiyar Mousavi-Sadr\authormark{1,2}, Mohammad H. Zhoolideh Haghighi\authormark{1,3}, and Fatemeh S. Tabatabaei\authormark{1}}

\address{\authormark{1}School of Astronomy, Institute for Research in Fundamental Sciences (IPM), P. O. Box 19395-5531, Tehran, Iran\\
\authormark{2}Iranian National Observatory (INO), Institute for Research in Fundamental Sciences (IPM), P. O. Box 19568-36613, Tehran, Iran\\
\authormark{3}Department of Physics, K. N. Toosi University of Technology, P. O. Box 15875-4416, Tehran, Iran}

\email{\authormark{*} email: rkarimi@ipm.ir} 


%
\begin{abstract} The discovery of exoplanets has expanded our understanding of planetary systems and opened new avenues for astronomical research. In this study, we present a machine learning (ML) framework for exoplanet identification using a time-series photometric dataset from the Kepler Space Telescope, comprising 3,198 flux measurements across 5,074 stars. We investigate the performance of four supervised classification algorithms, namely Random Forest, k-Nearest Neighbors (KNN), Decision Tree, and Logistic Regression, using a comprehensive set of evaluation metrics such as accuracy, precision, recall, F1-score, Area Under the Receiver Operating Characteristic Curve (AUC-ROC), confusion matrices, and learning curves. Among the models, Random Forest achieves the highest accuracy (99.8\%) and near-perfect F1-scores, demonstrating superior generalization and robustness. KNN also performs strongly, achieving 99.3\% accuracy, while Decision Tree demonstrates moderate performance with 97.1\% accuracy, and Logistic Regression trails behind with the lowest accuracy and generalization at 95.8\%. Notably, the application of the Synthetic Minority Over-sampling Technique (SMOTE) significantly improves performance across all models by addressing class imbalance. These findings underscore the effectiveness of ensemble-based machine learning techniques, particularly Random Forest, in handling large volumes of photometric data for automated exoplanet detection. This approach holds significant potential for implementation at ground-based facilities, such as the Iranian National Observatory (INO), where such extensive and precise datasets can further advance exoplanet discovery and characterization efforts.
\end{abstract}

\section{Introduction}
Exoplanets are planets located beyond our solar system. Since the pioneering discoveries of the first exoplanets \cite{Wolszczan}, \cite{Mayor}, the field of exoplanetary science has developed into a prominent and rapidly advancing branch of astrophysics \cite{Malik}. Planets are cooler, smaller, and far less luminous than their host stars, necessitating high-precision observational methods for their detection. To date, nearly 6,000 exoplanets have been discovered using a variety of detection methods, such as the transit \cite{Charbonneau}, radial velocity \cite{Campbell}, gravitational microlensing \cite{Mao}, direct imaging \cite{Chauvin}, and astrometry \cite{Benedict}.

The transit method is the most effective detection technique, responsible for discovering the majority of known exoplanets. This method detects periodic dips in a star's brightness caused by an orbiting planet passing in front of the stellar disk, temporarily blocking a portion of the emitted light. However, several factors can complicate the transit detection process. A primary limitation is that transits are only observable when a planet's orbital plane is nearly edge-on from the observer’s line of sight. Furthermore, the method is prone to a high false positive rate due to phenomena such as eclipsing binary stars, stellar variability, and starspots \cite{Winn}. Other detection methods have also played important roles, with over 1,000 exoplanets confirmed using radial velocity and several hundred identified through the techniques of gravitational microlensing, direct imaging, and astrometry.

Two main space telescope surveys for exoplanet detection are Kepler and the Transiting Exoplanet Survey Satellite (TESS). The Kepler Space Telescope \cite{Borucki}, launched in 2009, was NASA’s first mission dedicated to the discovery of exoplanets. Its primary objective was to search a specific region of the Milky Way, particularly within the constellations Cygnus and Lyra, for Earth-sized planets located in or near the habitable zones of Sun-like stars. This was accomplished through photometric monitoring of approximately 156,000 stars within its $115~\text{deg}^2$ field of view (FoV). Kepler’s second extended mission, known as K2 \cite{Howell}, continued the search for exoplanets, expanding the survey to include a broader and more diverse range of target stars. In 2018, the Transiting Exoplanet Survey Satellite (TESS) \cite{Ricker} was launched by NASA, designed to discover hundreds of transiting exoplanets smaller than Neptune. TESS is an ongoing mission that focuses on bright stars, which are well-suited for follow-up spectroscopic observations aimed at determining the planets’ masses and atmospheric compositions. Together, these missions, along with other space- and ground-based observatories such as the James Webb Space Telescope and the High Accuracy Radial velocity Planet Searcher (HARPS), represent a significant advancement in revolutionizing the field of exoplanet research.

Despite these advancements and the capabilities of modern instruments, many exoplanet candidates remain unconfirmed. The Kepler Space Telescope, K2, and TESS have detected thousands of potential exoplanets that require further detailed observations and analysis to determine whether they are genuine exoplanets or false positives \cite{Mahdiyar}. Traditionally, the classification of such signals has relied heavily on human expertise. Although trained individuals can effectively distinguish between real planetary signals and false detections, this approach presents two key limitations. First, large-scale transit surveys such as Kepler and TESS generate extensive light curve datasets, making manual analysis increasingly time-consuming and impractical as the volume of astronomical data continues to grow. Second, human judgment is inherently inconsistent, with factors such as cognitive bias contributing to variations in classification, even when the same observer evaluates identical signals. This inconsistency poses a challenge to accurately quantifying and correcting biases within the classification process. To overcome these challenges, developing a more efficient, accurate, and consistent system is essential for improving the reliability of exoplanet identification \cite{Tey}. This is where machine learning (ML) algorithms come into play, offering a powerful tool for the rapid identification, classification, and even prediction of planet-hosting systems \cite{Shallue, MousaviSadr, Valizadegan, Davoult, MousaviSadr2024}. In addition, machine learning has significantly advanced the analysis of complex datasets in multiple branches of astrophysics, as demonstrated in studies such as \cite{Zhoolideh, Zhoolideh2025, Zhoolideh20252, Richards2011, Moeller2020, Ayubinia2025}.

In this study, we utilize a dataset of Kepler light curves to evaluate and compare the performance of four widely used machine learning algorithms, including Logistic Regression, Decision Tree, Random Forest, and k-Nearest Neighbors, in the task of exoplanet detection. These algorithms are employed on Kepler’s time-series transit data, which contain slight fluctuations in stellar brightness due to possible planetary transits. By assessing each algorithm's performance based on classification accuracy, we seek to understand their effectiveness in handling large-scale exoplanet transit data. This comparison highlights the advantages and drawbacks of each method in the context of automated exoplanet detection.

This paper is organized as follows: Section 2 introduces the dataset and details the methodology, including data preprocessing, the implementation of machine learning models, and evaluation metrics. Section 3 presents the results and discusses the performance of each algorithm. Finally, Section 4 summarizes the findings and offers conclusions regarding the applicability of these methods to exoplanet identification.
\section{Data and methods}
\subsection{Dataset}
\begin{figure}[H]
 \centerline{\includegraphics[width=12cm]{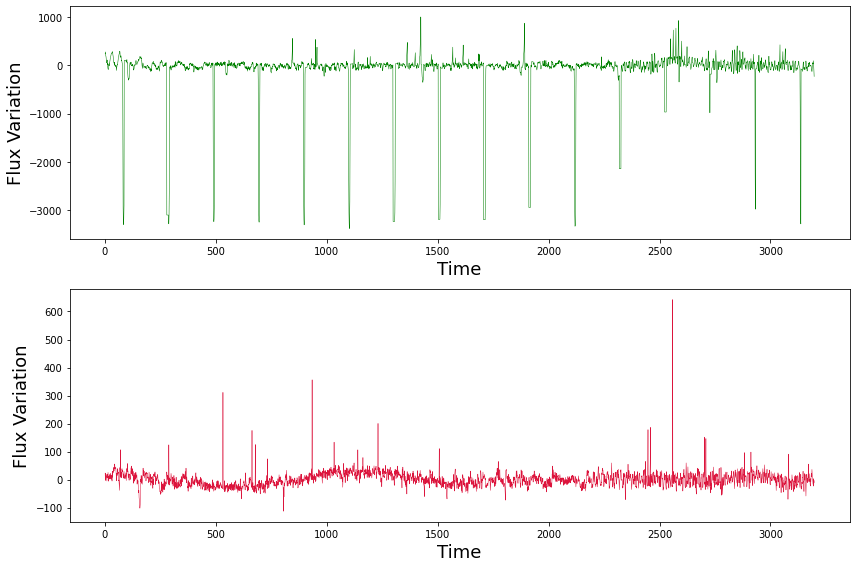}}
 \caption[]{Comparison of Kepler time-series light curves. Top: A star hosting an exoplanet exhibits periodic dips caused by transits. Bottom: A star without an exoplanet shows no such variations.}
 \label{fig1}
\end{figure}

In this study, we use time-series photometric observations of 5,074 stars from a publicly available dataset\footnote{\url{https://www.kaggle.com/datasets/keplersmachines/kepler-labelled-time-series-data}} from the Kepler Space Telescope mission, among which 37 have been confirmed to host exoplanets. The dataset contains 3,198 flux measurements for each star, representing variations in stellar brightness over time (see Figure~\ref{fig1}). These flux values, derived from raw photometric data collected by Kepler’s Charge-Coupled Device (CCD) detectors, serve as input features for the analysis \cite{Jenkins}.

The dataset includes labeled classifications indicating whether each light curve corresponds to an exoplanet-hosting star. Each star in the dataset is assigned a binary label: a value of 1 denotes that no exoplanet has been detected in the system, while a value of 2 indicates that the star has been confirmed to host at least one exoplanet. Such labeled data are crucial for training and evaluating machine learning models aimed at exoplanet identification. They serve as ground truth references, enabling the development of supervised learning algorithms capable of distinguishing planetary transits from other astrophysical phenomena, such as stellar variability, eclipsing binaries, and instrumental noise \cite{Shallue}.
\subsection{Data preprocessing}
Preprocessing steps are applied to ensure high-quality input for the machine learning models. Initially, the dataset is examined for missing values to confirm its completeness. Then, the class labels are standardized: a star without an exoplanet is assigned the value 0, while a star with an exoplanet is assigned the value 1.

Moreover, feature scaling, an important preprocessing technique in machine learning, is applied to the dataset. It transforms numerical features to a common scale to prevent those with larger ranges from disproportionately influencing the model’s performance. The stellar flux values in the dataset are normalized around zero and long-term variations, such as those resulting from stellar activity or instrumental effects, are removed from the light curves. This is a standard step in time-series analysis for exoplanet detection, as it eliminates baseline trends and enhances the visibility of short-term variations like transits.

Due to the imbalance between the number of exoplanet-hosting stars and stars without confirmed exoplanets in the dataset, training an ML model on this dataset could be challenging, as the learning algorithm might tend to favor the majority class \cite{Chawla}. To address this issue, the Synthetic Minority Over-sampling Technique (SMOTE) is applied to balance the class distribution. SMOTE creates synthetic samples of the minority class by interpolating between existing samples and their nearest neighbors. For a given minority class sample $X_i$, a synthetic sample $X_{\text{new}}$ is generated using:
\begin{equation}
X_{\text{new}}=X_i+\delta\cdot(X_{zi}-X_i),
\end{equation}
where $X_{zi}$ is one of the $k$-nearest neighbors of $X_i$, and $\delta$ is a random number between 0 and 1. This method generates new, plausible samples instead of merely duplicating existing ones, thereby reducing the risk of overfitting and enhancing the model’s generalization capability \cite{Chawla2002}. The class distribution is evaluated both before and after applying SMOTE, confirming that the technique effectively balances the dataset and improves the model’s ability to learn from both classes more equitably.

Finally, the dataset is divided into training and test sets using a 70:30 split ratio to help ensure an unbiased assessment of how well the models can identify exoplanet-hosting stars in new, real-world data. The training set, comprising 70\% of the data, is used to train and optimize the ML models, allowing them to learn the potential underlying patterns and relationships within the light curve features. The remaining 30\% of the data is held out as a test set, which serves as an independent dataset for evaluating the models' final performance and generalization ability on unseen examples.
\subsection{Models}
We employ four machine learning algorithms, including Logistic Regression, Decision Tree, Random Forest, and k-Nearest Neighbors (KNN), to classify the light curves into two categories: stars that host exoplanets and those that do not. Logistic Regression is a statistical model commonly used for binary classification tasks, appreciated for its simplicity and effectiveness, especially when the decision boundary is approximately linear. It applies the logistic (sigmoid) function to estimate the probability that a given input belongs to a particular class. Given an input feature vector $X$, the model computes the probability $P(Y=1\mid X)$ as follows:
\begin{equation}
P(Y=1|X)=\frac{1}{1+e^{-(\boldsymbol{w}\cdot\boldsymbol{X}+b)}},
\end{equation}
where $w$ represents the weights of the features, and $b$ is the bias term \cite{Daniel Jurafsky}.

Decision Tree is a non-parametric model that follows a tree-like structure for classification and regression tasks. It consists of consecutive nodes, where each node represents a condition on a feature in the dataset. The conditions take the form $X_j>X_{(j, \text{th})}$, where $X_j$ is the value of the feature at index $j$, and $X_{(j,\text{th})}$ is a threshold determined during the training process. The terminal nodes, or leaves, do not represent conditions but instead carry the assigned label for a particular path within the tree. Decision Trees offer several advantages, such as the ability to handle large datasets with many features and provide interpretability. However, they are susceptible to overfitting, particularly when the tree grows too deep and captures noise in the training data \cite{Baron}.

One of the most popular learning ensemble methods is Random Forest \cite{Breiman2001}, which combines multiple decision trees, each trained on a randomly selected subset of the original dataset. During training, random subsets of features are used to construct the conditions at individual nodes. This randomness reduces correlation between trees, leading to diverse tree structures. The final prediction of a Random Forest is obtained by aggregating individual tree predictions, typically through majority voting. This method is effective for datasets with a large number of features and provides improved accuracy compared to individual decision trees \cite{Baron}.

The k-Nearest Neighbors algorithm is a straightforward yet effective non-parametric classification method that assigns a class label to a new sample based on the majority class among its $k$ closest neighbors in the feature space. This algorithm stores all training examples $(x, f(x))$ in memory, where $x$ is defined as an $n$-dimensional feature vector like $(a_1, a_2, a_3,..., a_n)$, and $f(x)$ is the corresponding output label. To classify a query point $x_q$, kNN identifies the $k$ nearest training examples based on the Euclidean distance, calculated as:
\begin{equation}
D(x_q, x_i)=\sqrt{\sum_{r=1}^n\Bigl(a_r(x_q)-a_r(x_i)\Bigl)^2},
\end{equation}
where $a_r$ is the value of the $r^{th}$ feature. The final class assignment is made through a majority vote among the $k$ nearest neighbors \cite{Li}.

All machine learning classification models used in this study are implemented using the widely adopted \texttt{scikit-learn} library\footnote{\url{https://www.scikit-learn.org}}. This Python-based library provides a robust and consistent interface for building, training, and evaluating machine learning models, making it an ideal choice for reproducible and efficient model development.
\subsection{Evaluations}
We use four possible prediction outcomes, including True Positive (TP), True Negative (TN), False Positive (FP), and False Negative (FN), to evaluate the performance of each model. These outcomes form the foundation for calculating key evaluation metrics, offering a comprehensive view of each model's classification effectiveness. A True Positive occurs when the model correctly predicts the presence of an exoplanet. A True Negative happens when the model predicts the absence of an exoplanet, and there is indeed no exoplanet. A False Positive refers to a scenario in which the model incorrectly predicts the presence of an exoplanet when the star does not actually host one. Conversely, a False Negative occurs when the model fails to detect an existing exoplanet. From these values, essential evaluation metrics, namely accuracy, precision, recall, and F1-score, are computed, offering a well-rounded assessment of the model’s overall effectiveness and its ability to identify exoplanet-hosting stars. Table~\ref{table1} summarizes the evaluation metrics used to assess the performance of the classification models, along with their purpose and corresponding formulas.

To assess the model's ability to distinguish between classes, the Receiver Operating Characteristic (ROC) curve is employed, and the Area Under the Curve (AUC) score is computed \cite{Baron}. The ROC curve illustrates the trade-off between the Recall (see the third row in Table~\ref{table1}) and the Precision (see the second row in Table~\ref{table1}) across a range of classification thresholds. The AUC score provides a single-value summary of this performance, where higher values reflect better ability to distinguish between two classes \cite{Vujovic}. Additionally, we include the confusion matrix in our model evaluation. In this matrix, each row and column corresponds to a specific class, with the matrix values indicating the number of correctly and incorrectly classified samples. Ideally, perfect classification produces a diagonal matrix, where only the diagonal elements are non-zero, reflecting accurate predictions across all classes \cite{Baron}.

To further assess the generalization performance of the models and identify potential overfitting/underfitting, we plot learning curves based on the F1-score for each classifier. These curves illustrate how both the training and cross-validation scores change as the training set size increases. A small gap between the two curves generally indicates good generalization, while a large gap suggests overfitting. Conversely, underfitting is indicated when both training and validation scores are low and closely aligned, reflecting a model that is too simple to capture the underlying data patterns.

\begin{figure}[H] 
 \centerline{\includegraphics[width=13.5cm, height= 2.5cm]{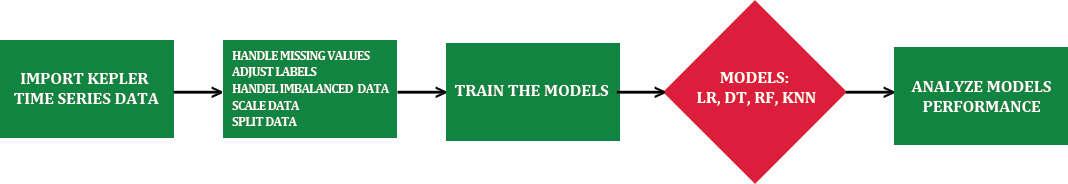}}
 \caption[]{An overview of our classification pipeline, encompassing key stages such as data preprocessing, training multiple machine learning models, and evaluation of their predictive performance. Data preprocessing involves several crucial steps, including checking for missing values, standardizing class labels, applying feature scaling, addressing class imbalance, and partitioning the dataset into training and testing subsets. The applied machine learning classification models are Logistic Regression (LR), Decision Tree (DT), Random Forest (RF), and k-Nearest Neighbors (KNN).}
 \label{fig2}
\end{figure}

Finally, the best-performing models are selected based on their comprehensive performance across all evaluation metrics, ensuring robust and reliable classification of exoplanet candidates. Figure~\ref{fig2} illustrates an overview of our classification pipeline, which includes essential stages such as data preprocessing, the training of multiple machine learning models, and comprehensive evaluation of their predictive capabilities. This structured approach allows for a systematic comparison of different algorithms and helps identify the most suitable model for accurate exoplanet detection.

\begin{table}[H]
\centering
\caption{Evaluation metrics used to assess the performance of the classification models, along with their respective purposes and formulas.}
\label{table1}
\begin{tabular}{@{} c c c @{}}
\toprule
\textbf{Metric Name} & \textbf{Purpose} & \textbf{Formula} \\
\midrule
\makecell{Accuracy} 
& \makecell{Measures the overall correctness}
& $\dfrac{TP + TN}{TP + TN + FP + FN}$ \\
\hline
\makecell{Precision}
& \makecell{Proportion of correctly predicted exoplanets \\among all predicted exoplanets}
& $\dfrac{TP}{TP + FP}$ \\
\hline
\makecell{Recall}
& \makecell{Proportion of correctly predicted exoplanets \\among all actual exoplanets}
& $\dfrac{TP}{TP + FN}$ \\
\hline
\makecell{F1-Score}
& \makecell{Harmonic mean of Precision and Recall}
& $2 \times \Biggl(\dfrac{\text{Precision} \times \text{Recall}}{\text{Precision} + \text{Recall}}\Biggl)$ \\
\bottomrule
\end{tabular}
\end{table}

\section{Results}
After applying the preprocessing techniques, particularly the SMOTE method, the resulting dataset comprises 10,074 stars, evenly split between exoplanet-hosting and non-exoplanet-hosting stars. The light curves in this balanced dataset contain no missing values, have properly adjusted labels, and feature values that have been scaled for consistency. The machine learning classification models, including Logistic Regression, Decision Tree, Random Forest, and k-Nearest Neighbors (KNN), are applied to this newly processed dataset to evaluate their performance in distinguishing exoplanet-hosting stars from those that do not host exoplanets.

Table~\ref{table2} summarizes the classification performance of the four machine learning models, reporting key evaluation metrics such as precision, recall, F1-score, accuracy, and the Area Under the ROC Curve (AUC). The results show that the Random Forest model outperforms the other classifiers, achieving the highest accuracy of 99.8\% and an outstanding AUC close to 1. It also demonstrates perfect recall for both exoplanet-hosting and non-exoplanet-hosting stars, with F1-scores reaching 0.998. These findings indicate that Random Forest is a highly robust and reliable model, making it particularly well-suited for the task of exoplanet classification. The k-Nearest Neighbors model also demonstrates strong performance, achieving an accuracy of 99.3\% and an AUC of 0.994. It attains F1-scores of 0.993 for both exoplanet-hosting stars and non-exoplanet-hosting stars, indicating a high degree of classification reliability. The Decision Tree model achieves an accuracy of 97.1\% and an AUC of 0.976. It maintains solid F1-scores, 0.972 for exoplanet-hosting stars and 0.971 for non-exoplanet-hosting stars, and its overall performance remains reasonably strong. Logistic Regression exhibits the weakest performance among the four models, with an accuracy of 95.8\% and an AUC of 0.969. It yields F1-scores of 0.959 for the exoplanet class and 0.958 for the non-exoplanet class. While its performance is still acceptable, the relatively lower recall values indicate that Logistic Regression may have difficulty capturing more complex patterns in the data, rendering it less suitable for high-precision exoplanet classification in this context.

\begin{table}[H]
\centering
\caption{Classification performance metrics of the four machine learning models in distinguishing between exoplanet-hosting and non-exoplanet-hosting stars. The reported metrics are precision, recall, and F1-score for each class, as well as overall accuracy and Area Under the ROC Curve (AUC).}
\resizebox{\textwidth}{!}{%
\begin{tabular}{|l|ccc|ccc|c|c|}
\hline
\textbf{Model} & \multicolumn{3}{c|}{\textbf{Exoplanet}} & \multicolumn{3}{c|}{\textbf{Non-Exoplanet}} & \textbf{Accuracy} & \textbf{AUC} \\
\cline{2-7}
& \textbf{Precision} & \textbf{Recall} & \textbf{F1-score} & \textbf{Precision} & \textbf{Recall} & \textbf{F1-score} & & \\
\hline
Random Forest & 0.996 & 1.0 & 0.998 & 1.0 & 0.996 & 0.998 & 99.8\% & 0.999 \\
k-Nearest Neighbors & 0.987 & 1.0 & 0.993 & 1.0& 0.987 & 0.993 & 99.3\% & 0.994 \\
Decision Tree & 0.954 & 0.990 & 0.972 & 0.989 & 0.953 & 0.971 & 97.1\% & 0.976 \\
Logistic Regression & 0.940 & 0.979 & 0.959 & 0.978 & 0.938 & 0.958 & 95.8\% & 0.969 \\
\hline
\end{tabular}
}
\label{table2}
\end{table}

\begin{figure}[H]
    \centering
    \begin{subfigure}[b]{0.45\textwidth}
        \includegraphics[width=\textwidth, height=5.5cm]{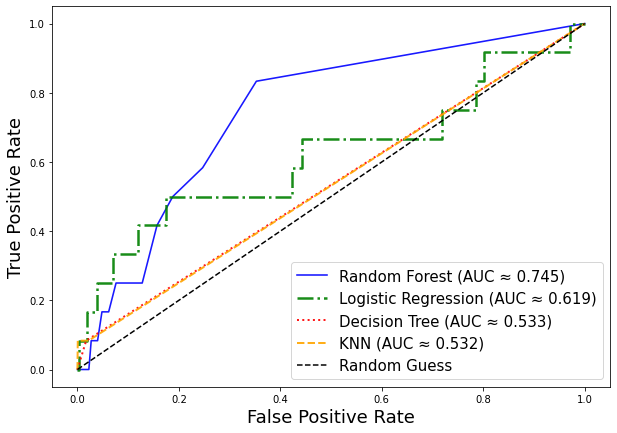}
    \end{subfigure}
    \hfill
    \begin{subfigure}[b]{0.45\textwidth}
        \includegraphics[width=\textwidth, height=5.5cm]{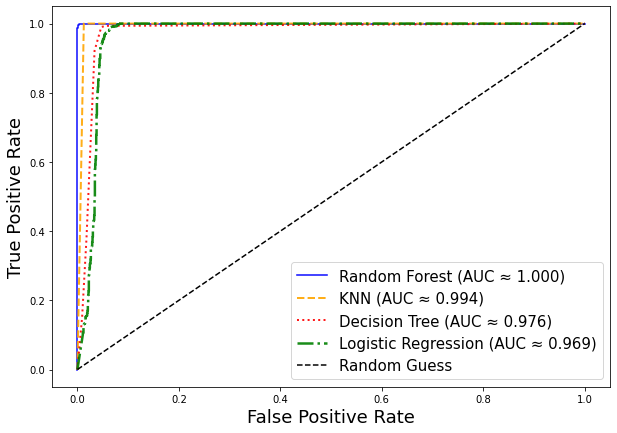}
    \end{subfigure}
    \caption{ROC curves for all models before (left) and after (right) applying SMOTE. These plots illustrate a substantial improvement in classification performance across all models following the oversampling process. Notably, the Random Forest model achieved a near-perfect AUC of $\sim$1, highlighting the effectiveness of SMOTE in addressing class imbalance and enhancing the models' discriminative power.}
    \label{fig3}
\end{figure}

Figure~\ref{fig3} illustrates the comparison of ROC curves for all models before and after applying the SMOTE oversampling method. A random guess (the black dashed line) corresponds to the diagonal running from the bottom-left to the top-right of the ROC space, where the true positive rate equals the false positive rate, indicating no ability to distinguish between classes. A model’s discriminatory power is reflected by how much its ROC curve deviates above this diagonal. The closer the curve is to the top-left corner and the farther it is from the random guess line, the better the model’s classification performance. An ideal model’s ROC curve is located near the top-left corner of the plot, indicating a high true positive rate and thus a strong ability to correctly identify planet-hosting stars. Prior to oversampling, the Random Forest model achieved the highest AUC score of 0.745, outperforming Logistic Regression (0.619), Decision Tree (0.533), and KNN (0.532). After applying SMOTE, the performance of all models improved substantially. In particular, the Random Forest model attained a near-perfect AUC of $\sim$1, followed closely by KNN (0.994), Decision Tree (0.976), and Logistic Regression (0.969). These results highlight the effectiveness of SMOTE in mitigating class imbalance and significantly enhancing the discriminatory power of the models.

\begin{figure}
    \centering
    \begin{subfigure}[b]{0.45\textwidth}
        \includegraphics[width=\textwidth, height=6cm]{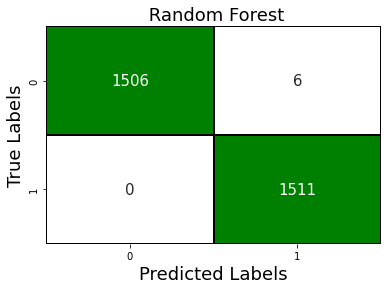}
    \end{subfigure}
    \hfill
    \begin{subfigure}[b]{0.45\textwidth}
        \includegraphics[width=\textwidth, height=6cm]{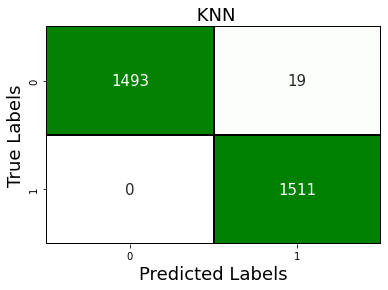}
    \end{subfigure}

    \vspace{0.5cm}

    \begin{subfigure}[b]{0.45\textwidth}
        \includegraphics[width=\textwidth, height=6cm]{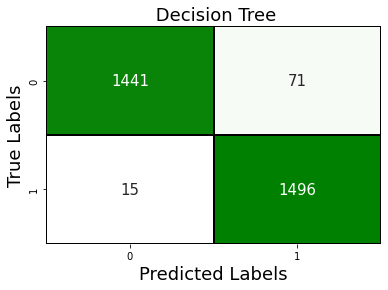}
    \end{subfigure}
    \hfill
    \begin{subfigure}[b]{0.45\textwidth}
        \includegraphics[width=\textwidth, height=6cm]{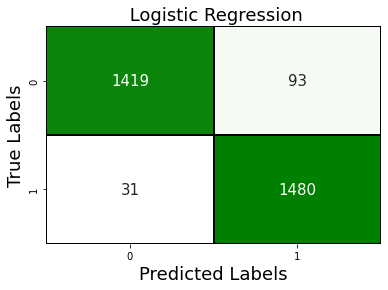}
    \end{subfigure}
    \caption{Confusion matrices for our four classification models. Random Forest achieves near-perfect classification with minimal errors, closely followed by KNN. The Decision Tree model shows increased misclassification rates, especially in false positives, while Logistic Regression exhibits the highest number of both false positives and false negatives, reflecting comparatively weaker performance.}
    \label{fig4}
\end{figure}

The confusion matrices in Figure~\ref{fig4} provide additional insight into the performance of the classification models by detailing their predictions in terms of true positives, true negatives, false positives, and false negatives. These matrices visually represent how well each model distinguishes between the two classes. In this context, class 0 represents stars that do not host exoplanets, while class 1 denotes stars that do. The Random Forest model demonstrates near-perfect classification performance, producing only 6 false positives and no false negatives. It accurately identifies 1,506 non-exoplanet-hosting stars and correctly classifies all 1,511 exoplanet-hosting stars. These results underscore the model’s exceptional accuracy, precision, and recall, highlighting its robustness and reliability in distinguishing between the two classes with minimal misclassification. The k-Nearest Neighbors model also exhibits strong performance, misclassifying only 19 non-exoplanet-hosting stars as exoplanet-hosting and achieving perfect recall for the class 1. With 1,493 true negatives and 1,511 true positives, the model demonstrates high sensitivity and precision. However, the slightly higher number of false positives compared to the Random Forest model indicates a minor trade-off in specificity, suggesting that while KNN is highly effective, it is marginally less conservative in distinguishing non-exoplanet cases. The Decision Tree model shows a moderate decrease in classification performance, with 71 false positives and 15 false negatives. Despite correctly identifying 1,441 non-exoplanet-hosting stars and 1,496 exoplanet-hosting stars, the increased number of misclassifications suggests a diminished generalization capability relative to the Random Forest and k-Nearest Neighbors models. This outcome aligns with its comparatively lower overall accuracy and AUC, underscoring its relative limitations in classification performance. Logistic Regression exhibits the highest misclassification rate among all evaluated models, with 93 false positives and 31 false negatives. It correctly identifies 1,419 non-exoplanet-hosting stars and 1,480 exoplanet-hosting stars. These outcomes highlight the model’s limited ability to effectively distinguish between the two classes, which aligns with its lower precision, recall, and F1-scores. While the model remains functional, its relatively high error rates suggest that it is less well-suited for applications demanding high classification accuracy and robustness.

Overall, the confusion matrices confirm that the Random Forest and KNN models deliver the most reliable classification results, demonstrating minimal misclassifications. In contrast, the Decision Tree and Logistic Regression models exhibit higher error rates, particularly in accurately identifying non-exoplanet-hosting stars.

\begin{figure}
    \centering
    \begin{subfigure}[b]{0.45\textwidth}
        \includegraphics[width=\textwidth, height=5.5cm]{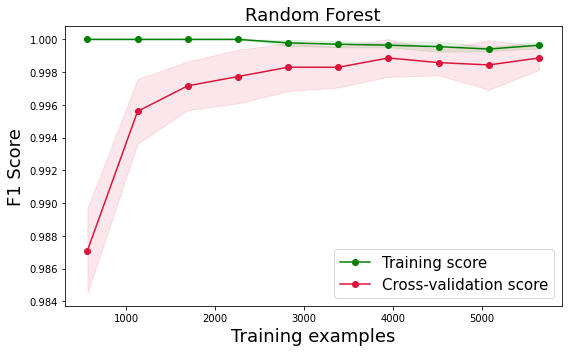}
    \end{subfigure}
    \hfill
    \begin{subfigure}[b]{0.45\textwidth}
        \includegraphics[width=\textwidth, height=5.5cm]{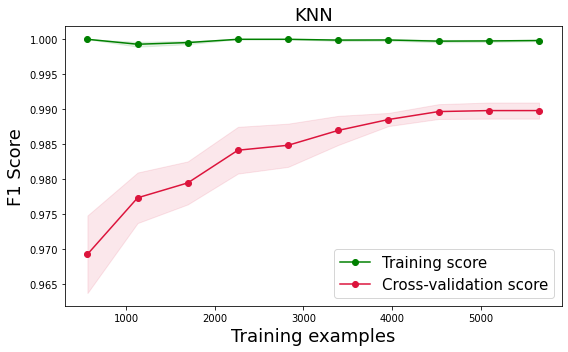}
    \end{subfigure}

    \vspace{0.5cm}

    \begin{subfigure}[b]{0.45\textwidth}
        \includegraphics[width=\textwidth, height=5.5cm]{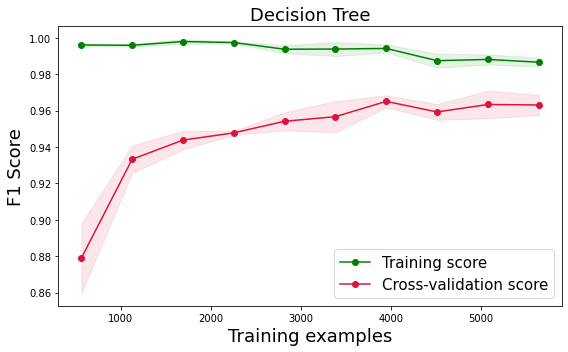}
    \end{subfigure}
    \hfill
    \begin{subfigure}[b]{0.45\textwidth}
        \includegraphics[width=\textwidth, height=5.5cm]{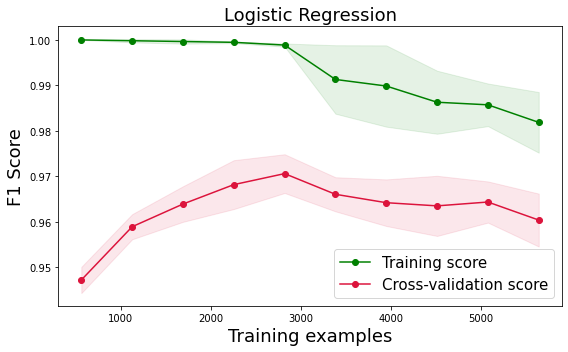}
    \end{subfigure}
    \caption{Learning curves of our four classification models. Displaying the F1-score as a function of the number of training examples. The green line indicates the training score, while the red line represents the cross-validation score. Shaded areas around each curve illustrate the standard deviation.}
    \label{fig5}
\end{figure}

The learning curves in Figure~\ref{fig5} illustrate the F1-score performance of the four classification models as a function of the training set size, offering insights into their learning behavior and generalization capabilities. The Random Forest model demonstrates outstanding learning dynamics, where both training and cross-validation F1-scores stay consistently high across all training set sizes. The cross-validation score rapidly converges to above 0.997, indicating strong generalization. Additionally, the narrow gap between the training and cross-validation curves reflects low variance and stable performance. The k-Nearest Neighbors model shows a steady improvement in cross-validation performance as the training size increases. The training score remains perfect throughout, while the cross-validation score rises from approximately 0.970 to 0.990. The narrowing gap between the two curves implies that the model benefits from more data and becomes increasingly generalizable. However, the consistently perfect training score may suggest slight overfitting, particularly at lower training volumes. The Decision Tree model demonstrates a more pronounced gap between the training and cross-validation curves. While the training score remains very high (near 0.999), the cross-validation score starts lower and gradually improves, reaching around 0.965. This wider gap, along with variability in the validation curve, indicates signs of overfitting, evident from the near-perfect training performance contrasted with the lower validation performance. However, as the training set grows, the gap narrows, suggesting that the model's generalization ability improves with more data. For logistic regression, as the training size increases, the F1 score on the training set gradually declines from 1.0 to approximately 0.980, a typical and desirable pattern reflecting reduced overfitting. Concurrently, the cross-validation F1 score begins around 0.950 and stabilizes near 0.960 after initial fluctuations, indicating that the model benefits from increased training data, which helps lower variance and improve generalization. The relatively narrow gap between training and validation scores, along with consistently high F1 values, suggests a favorable bias-variance tradeoff, implying that the model is neither underfitting nor significantly overfitting. Nevertheless, the learning curves of our other models exhibit even more promising performance for distinguishing exoplanets from non-exoplanets.

In general, the learning curves reinforce that Random Forest is the most stable and effective model for this classification task, consistently achieving high performance with minimal variance. KNN follows closely, benefiting from increased data. The Decision Tree model exhibits improvement with larger training sets but remains prone to slightly overfitting.  Logistic Regression, while stable and consistent, shows signs of reduced overfitting due to its linear nature, limiting its ability to model the complexity of exoplanet classification.

\section{Summary and conclusion}
We analyze a dataset obtained from the Kepler Space Telescope mission, comprising time-series photometric observations of 5,074 stars, to assess the effectiveness of four supervised classification algorithms, including Random Forest, k-Nearest Neighbors (KNN), Decision Tree, and Logistic Regression, in identifying exoplanet-hosting stars. Preprocessing steps are applied to this dataset to ensure high-quality input for the machine learning models. Initially, the dataset is checked for missing values to verify its completeness. To improve the clarity of transit signals and reduce noise, the stellar flux values are normalized around zero, and long-term variations, often caused by stellar activity or instrumental effects, are removed from the light curves. Notably, to address the imbalance between the number of exoplanet-hosting and non-exoplanet-hosting stars, the Synthetic Minority Over-sampling Technique (SMOTE) is employed to balance the class distribution. This preprocessing step leads to a substantial improvement in the performance of all classification models.

The classification models are evaluated using a comprehensive set of performance metrics, including accuracy, precision, recall, F1-score, AUC-ROC, confusion matrices, and learning curves. The results indicate that the Random Forest model consistently outperforms all other models across all evaluation criteria. It achieves the highest accuracy of 99.8\%, an outstanding AUC of 0.999, and nearly perfect F1-scores for both exoplanet and non-exoplanet classes. Its confusion matrix reveals only 6 misclassified instances, and its learning curve demonstrates excellent generalization. The KNN model also performs well, achieving an accuracy of 99.3\% and an AUC of 0.994. Although it exhibits slightly more false positives than the Random Forest model, its overall precision and recall remain high. The learning curve for KNN indicates consistent improvement in performance with increasing training data, suggesting strong scalability and generalization capability. The Decision Tree model, while achieving reasonably high performance (97.1\% accuracy and an AUC of 0.976), shows greater susceptibility to overfitting. This is evidenced by a noticeable gap between training and cross-validation scores and higher misclassification rates in the confusion matrix. These factors suggest that, although Decision Tree is a viable classification option, it is less stable and less accurate compared to Random Forest and KNN.  Logistic Regression demonstrates the weakest performance among the evaluated models, with an accuracy of 95.8\% and an AUC of 0.969. The model shows signs of reduced overfitting, as reflected in the learning curves and confusion matrix, while it can serve as a reliable baseline model. Its linear structure limits its capacity to capture the complex patterns present in exoplanet time-series data. As a result, Logistic Regression is less suitable for this classification task compared to the other three models.

In conclusion, Random Forest emerges as the most suitable algorithm for identifying exoplanets from Kepler time-series data, offering exceptional performance, generalization, and stability across all evaluation metrics. These findings, consistent with those reported by \cite{McCauliff}, reinforce that ensemble-based models, particularly Random Forest, are among the most effective approaches for exoplanet identification tasks. Meanwhile, the KNN model serves as a strong alternative, especially when simplicity and ease of implementation are prioritized, while still maintaining high accuracy and reliable classification performance.

These machine learning models, particularly Random Forest and k-Nearest Neighbors, hold significant promise for application at the Iranian National Observatory (INO) as it initiates its search for planets beyond our solar system. By leveraging photometric data collected from INO’s world-class instruments, most notably its 3.4-meter optical telescope, these models can efficiently and accurately detect the subtle transit signals that indicate the presence of exoplanets. Their strong generalization capabilities position them as powerful tools in supporting INO’s future contributions to the global effort of discovering and characterizing exoplanets.

\section{Acknowledgments}
The authors acknowledge the usage of the \texttt{Scikit-learn} \cite{Pedregosa}, \texttt{matplotlib} \cite{Hunter}, and \texttt{NumPy} \cite{VanDerWalt} libraries.

%

\end{document}